\documentstyle[psfig]{mn}

\newif\ifAMStwofonts
\ifoldfss
  \ifCUPmtlplainloaded \else
    \NewTextAlphabet{textbfit} {cmbxti10} {}
    \NewTextAlphabet{textbfss} {cmssbx10} {}
    \NewMathAlphabet{mathbfit} {cmbxti10} {} % for math mode
    \NewMathAlphabet{mathbfss} {cmssbx10} {} %  "   "    "
  \fi
  \ifAMStwofonts
    \ifCUPmtlplainloaded \else
      \NewSymbolFont{upmath} {eurm10}
      \NewSymbolFont{AMSa} {msam10}
      \NewMathSymbol{\upi}     {0}{upmath}{19}
      \NewMathSymbol{\umu}     {0}{upmath}{16}
      \NewMathSymbol{\upartial}{0}{upmath}{40}
      \NewMathSymbol{\leqslant}{3}{AMSa}{36}
      \NewMathSymbol{\geqslant}{3}{AMSa}{3E}

    \fi
  \fi
\fi % End of OFSS

\ifnfssone
  \newmathalphabet{\mathit}
  \addtoversion{normal}{\mathit}{cmr}{m}{it}
  \addtoversion{bold}{\mathit}{cmr}{bx}{it}
  \newmathalphabet{\mathbfit} % math mode version of \textbfit{..}
  \addtoversion{normal}{\mathbfit}{cmr}{bx}{it}
  \addtoversion{bold}{\mathbfit}{cmr}{bx}{it}
  \newmathalphabet{\mathbfss} % math mode version of \textbfss{..}
  \addtoversion{normal}{\mathbfss}{cmss}{bx}{n}
  \addtoversion{bold}{\mathbfss}{cmss}{bx}{n}
  \ifAMStwofonts
    \ifCUPmtlplainloaded \else
      \UseAMStwoboldmath
      \makeatletter
      \new@mathgroup\upmath@group
      \define@mathgroup\mv@normal\upmath@group{eur}{m}{n}
      \define@mathgroup\mv@bold\upmath@group{eur}{b}{n}
      \edef\UPM{\hexnumber\upmath@group}
      \new@mathgroup\amsa@group
      \define@mathgroup\mv@normal\amsa@group{msa}{m}{n}
      \define@mathgroup\mv@bold\amsa@group{msa}{m}{n}
      \edef\AMSa{\hexnumber\amsa@group}
      \makeatother
      \mathchardef\upi="0\UPM19
      \mathchardef\umu="0\UPM16
      \mathchardef\upartial="0\UPM40
      \mathchardef\leqslant="3\AMSa36
      \mathchardef\geqslant="3\AMSa3E
    \fi
  \fi
\fi % End of NFSS release 1

\ifnfsstwo
  \DeclareMathAlphabet{\mathbfit}{OT1}{cmr}{bx}{it}
  \SetMathAlphabet\mathbfit{bold}{OT1}{cmr}{bx}{it}
  \DeclareMathAlphabet{\mathbfss}{OT1}{cmss}{bx}{n}
  \SetMathAlphabet\mathbfss{bold}{OT1}{cmss}{bx}{n}
  \ifAMStwofonts
    \ifCUPmtlplainloaded \else
      \DeclareSymbolFont{UPM}{U}{eur}{m}{n}
      \SetSymbolFont{UPM}{bold}{U}{eur}{b}{n}
      \DeclareSymbolFont{AMSa}{U}{msa}{m}{n}
      \DeclareMathSymbol{\upi}{0}{UPM}{"19}
      \DeclareMathSymbol{\umu}{0}{UPM}{"16}
      \DeclareMathSymbol{\upartial}{0}{UPM}{"40}
      \DeclareMathSymbol{\leqslant}{3}{AMSa}{"36}
      \DeclareMathSymbol{\geqslant}{3}{AMSa}{"3E}
    \fi
  \fi
\fi % End of NFSS release 2

\ifCUPmtlplainloaded \else
  \ifAMStwofonts \else % If no AMS fonts
    \def\upi{\pi}
    \def\umu{\mu}
    \def\upartial{\partial}
  \fi
\fi

\title[PSR J1410$-$6132] {PSR~J1410$-$6132: A young, energetic pulsar
associated with EGRET source 3EG J1410-6147}
\author[J. T. O'Brien et al.]{ J.~T.~O'Brien,$^1$\thanks{Email: Jennifer.O'Brien@manchester.ac.uk} 
S.~Johnston,$^2$ M.~Kramer,$^1$ A.~G.~Lyne,$^1$ M.~Bailes,$^3$ A.~Possenti,$^4$ 
\newauthor
M.~Burgay,$^4$ D.~R.~Lorimer,$^5$ M.~A.~McLaughlin,$^5$ G.~Hobbs,$^2$ 
D.~Parent$^6$ and \newauthor L.~Guillemot$^6$ \\
$^1$University of Manchester, Jodrell Bank Centre for Astrophysics, 
Alan-Turin Building, Manchester M13 9PL, UK \\
$^2$Australia Telescope National Facility, CSIRO, P.O.~Box~76, Epping
NSW~1710, Australia \\
$^3$Swinburne Centre for Astrophysics and Supercomputing, Swinburne 
University of Technology, Hawthorn, VIC 3122, Australia \\
$^4$INAF - Osservatorio Astronomica di Cagliari, 09012 Capoterra, Italy\\
$^5$Department of Physics, West Virginia University, Morgantown, WV 26506, USA \\
$^6$Centre d'Etudes Nucleaires de Bordeaux Gradignan, Chemin du Solarium, Le
Haut Vigneau, BP 120, F-33175 GRADIGNAN Cedex, France
}

\date{\today}

\pagerange{\pageref{firstpage}--\pageref{lastpage}}
\pubyear{2007}
\begin{document}
\maketitle
\label{firstpage}

\begin{abstract}
  We present the discovery of PSR~J1410$-$6132, a 50-ms pulsar found during a
  high-frequency survey of the Galactic plane, using a 7-beam 6.3-GHz receiver
  on the 64-m Parkes radio telescope.  The pulsar lies within the error box of
  the unidentified EGRET source 3EG J1410$-$6147, has a characteristic age of
  26 kyr and a spin-down energy of 10$^{37}$~erg s$^{-1}$. It has a very high
  dispersion measure of 960 $\pm10$ cm$^{-3}$ pc and the largest rotation
  measure of any pulsar, RM=$+2400 \pm30$ rad m$^{-2}$. The pulsar is very
  scatter-broadened at frequencies of 1.4 GHz and below, making pulsed
  emission almost impossible to detect. Assuming a distance of 15~kpc, the
  pulsar's spin-down energy and a $\gamma$-ray efficiency factor of $\sim$10
  per cent is sufficient to power the $\gamma$-ray source. We therefore
  believe we have identified the nature of 3EG J1410$-$6147. This new
  discovery suggests that deep targeted high-frequency surveys of inner-galaxy
  EGRET sources could uncover further young, energetic pulsars.
\end{abstract}


\section{Introduction}
\label{intro}

The Galactic population of pulsars is still poorly known in the inner parts of
our Galaxy. Studies of the Galactic structure (e.g.~Bahcall 1986, Gilmore et
al.~1989)\nocite{bah86,gwk89} and studies of the radial distribution of
pulsars (e.g.~Lyne et al.~1985)\nocite{lmt85} however, suggest that a large
number of pulsars await discovery in the inner spiral arms. Unfortunately,
selection effects imposed by the interstellar medium make the discovery of
such pulsars difficult. First, the interaction of the broadband pulsed radio
signal with the ionised component of the interstellar medium causes the effect
of dispersion, delaying signals observed at lower frequencies relative to
their higher frequency counterparts (Hewish et al.~1968)\nocite{hbp+68}.  The
dispersive delay is inversely proportional to the square of the observing
frequency, the constant of proportionality being the column density of the
free electrons between Earth and the pulsar, known as the dispersion measure
(DM). Without a correction for this effect, the recorded pulse becomes smeared
and eventually undetectable. Secondly, the pulse also becomes broadened due to
interstellar multi-path scattering as the free electron distribution in the
turbulent interstellar medium is inhomogeneous \cite{r77}.  Scattering is
especially severe at low frequencies, since the scattering time is
proportional to $\nu^{\beta}$ where $\nu$ is the observational frequency and
$\beta\ga-4$ (L\"ohmer et al.~2004)\nocite{lmg+04}.  Unlike dispersion,
scattering cannot be removed by instrumental means. Both effects are
particularly severe for high DM pulsars. Another factor that mitigates against
low frequency surveys is that at low Galactic latitudes, the background
(synchrotron) radiation has an effective temperature, T$_{sky}$, which
dominates T$_{sys}$ at low frequencies with a dependence which varies as
approximately $\nu^{-2.6}$.

Any pulsar survey therefore needs to balance the above effects against the
fact that pulsars are intrinsically brighter at low frequencies and the larger
telescope beam at lower frequencies reduces the survey time for a given area.
Population studies take these selection effects into account and a recent
population study by Lorimer et al.~(2006)\nocite{lfl+06} suggests that, using
the electron density model of Cordes \& Lazio (2002)\nocite{cl02}, a dearth of
pulsars exist in the inner Galaxy.  As uncertainties in the electron model
remain (cf.~Kramer et al.~2003, Lorimer et al.~2006)\nocite{kbm+03} a
high-frequency survey is the only direct way to shed light on the population
of pulsars in the inner Galaxy, as was previously demonstrated by the
pioneering use of high frequencies in pulsar surveys by Clifton \& Lyne (1986)
\nocite{cl86}. This is particularly true for the studies of young pulsars
which are especially important for our understanding of birthrates derived in
these population studies. Indeed, for young pulsars still residing near their
birth place in the Galactic plane, the aforementioned selection effects are
extremely severe as they are typically spinning fast. These considerations
were the motivation to conduct a survey of the inner Galactic plane at an
unusually high frequency of 6.3 GHz, of which the first results are presented
here.

The new survey utilises a seven beam receiver that was built in collaboration
between Jodrell Bank Observatory and the Australia Telescope National
Facility, operating at a wavelength around 5\,cm, to search the Galaxy for
emission of methanol masers. Exploiting this ``methanol multi-beam'' (MMB)
receiver (each beam with a width of 0.11 deg) allows us to rapidly cover the
inner Galactic plane at a central observing frequency of 6306\,MHz, with a
bandwidth of 576\,MHz, which is indeed much higher than for usual pulsar
surveys. As a result, we expect the survey to discover a sample of new pulsars
that is dominated by young objects.

We report here the first discovery of the MMB survey which, in accordance with
the expectations, is a young, energetic pulsar. In Section 2 we briefly
outline the parameters of the MMB survey and in Section 3 discuss the
pulsar's parameters and its discovery. Section 4 discuss the likely
association between the pulsar and the unidentified $\gamma$-ray
point source, 3EG J1410$-$6147.

\section{The Methanol Multibeam Survey}
\begin{figure}
\begin{center}
\centerline{\psfig{file=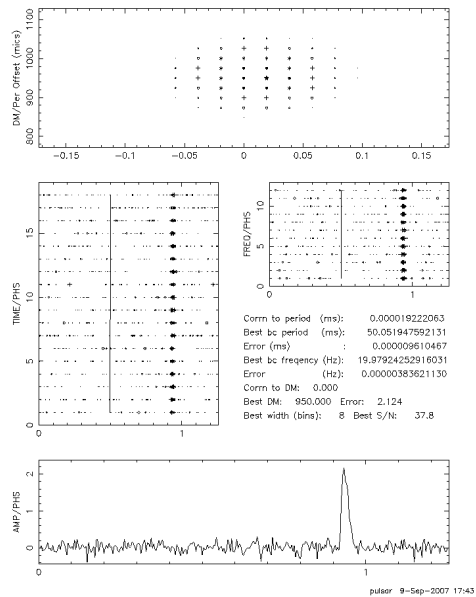,height=8cm} }
\end{center}
\caption{The confirmation observation of PSR J1410$-$6132. Top panel shows the
  search range in DM and period space, middle panels show the folded
  sub-integrations of the time series (left) and frequency bands (right),
  bottom panel shows the folded pulsar profile.}
\label{fig:profiles}
\end{figure}
The survey of the Galactic plane for pulsars using the MMB
receiver on the Parkes radio telescope commenced in February 2006.
Full details will be given elsewhere, therefore we will only summarise its basic characteristics now.

The survey used a \textbf{192-channel filterbank, with a channel width of 3~MHz for a total of 576\,MHz centred at} 6306\,MHz, \textbf{to cover} an area of 40
square degrees, ranging from a longitude of $-60\,^{\circ}<l<+20\,^{\circ}$
and a latitude range of $|b|<0.25\,^{\circ}$. The integration time of 1055~s
is modest and a compromise between survey speed and sensitivity. \textbf{Assuming a duty cycle of 5\% and with a system noise of 60 Jy , the average limiting detectable flux density is 0.1 mJy.} All data are
\textbf{1-bit} sampled at 0.125~ms, written to tape and analysed off-line using standard
algorithms which search for short- and long-period pulsars, as well as for
short radio bursts and bright single pulses (see Lorimer \& Kramer 2005 for
details).  Currently, 94\% of all survey pointings have been observed; their
full analysis will be presented elsewhere.

\section{PSR~J1410$-$6132}

Wide survey observations at the pulsar's location were carried out on 2007
April 18 and subsequent processing revealed a promising candidate with a pulse
period of 50~ms and a DM of
$\sim$800\,cm$^{-3}$\,pc with a signal$-$to$-$noise ratio of
12.6. Observations were made of the same area of sky using the same observing
setup on 2007 September 9 and the pulsations were confirmed (see
Figure~\ref{fig:profiles}).  In the following days, observations were made at
1.4 and 3.1~GHz using a digital backend capable of recording full Stokes
parameters. Further timing observations have been carried out since, and
additional polarimetric observations were made at 6.4~GHz.

The top part of Table~\ref{timing} gives the measured pulsar parameters
obtained from a coherent timing solution using 24 independent measurements
made over a time span of 98~days. The DM is measured from the time delay and
the RM obtained from the position angle variation with frequency across the
1~GHz band at 3.1~GHz.  Errors given in parentheses refer to the last quoted
digit and are twice the formal standard error.  The bottom part of the Table
gives the derived parameters where the distance has been computed from the DM
and the Cordes \& Lazio (2002) electron density model and the other parameters
using standard equations (Lorimer \& Kramer 2005). \nocite{lk05}

\begin{table}
\caption{Measured and derived parameters for PSR~J1410$-$6132}
\begin{center}
\begin{tabular}{ll}
\hline
Right Ascension (J2000)                 & $14^{\rm h}10^{\rm m}24^{\rm s}(4)$ \\
Declination (J2000)                     & $-61^{\circ}32\arcmin (1)$\\
Galactic Longitude (degrees)            & 312.19    \\
Galactic Latitude (degrees)             & --0.09     \\
Period (s)                              & 0.0500519462    \\
Period Derivative                       & $32\times 10^{-15}$ \\
Frequency (s$^{-1}$)                    & 19.979        \\
Frequency Derivative                    & $-1.267\times 10^{-11}$ \\ 
Epoch of Period (MJD)                   & 54357.3       \\
Dispersion Measure (cm$^{-3}$ pc)       & 960(10)       \\
Rotation Measure (rad m$^{-2}$)         & 2400(30) \\
Flux density at 1.5/3.0/6.5 GHz (mJy)   & \bf{6}(1)/2.0(2)/0.60(6) \\
\noalign{\medskip}
Spectral index                          & $-1.3(2)$ \\
Distance (kpc)                          & $15.3$     \\
Characteristic Age (yr)                 & $26\times 10^3$      \\
Surface Magnetic Field (G)              & $1.3\times 10^{12}$   \\
Spin down energy (erg s$^{-1}$)         & $1.0\times 10^{37}$    \\
\hline
\end{tabular}
\end{center}
\label{timing}
\end{table}

This pulsar is a young, highly energetic pulsar buried deep in the Galactic
plane. As a young object, it is the sixth shortest period non-recycled pulsar,
is one of only 12 pulsars with $\dot{E}>10^{37}$ergs$^{-1}$ and ranks in the
top 50 when sorted by age. Its DM, refined from the initial survey estimate,
is also very large, especially considering its displacement of 48$^{\circ}$
from the Galactic centre. Furthermore, the magnitude of the Rotation Measure
(RM) of $+2400\pm30$ rad m$^{-2}$ is also unusually large for this longitude,
and in fact, it is the largest for any known pulsar; nearly all other pulsars
in the vicinity have large negative RMs and extragalactic sources also have
large negative RMs in this area of sky \cite{bhg+07}.  The RM in conjunction
with the DM gives the mean value of the magnetic field parallel to the line of
sight as 3.1~$\mu$G.

\begin{figure}
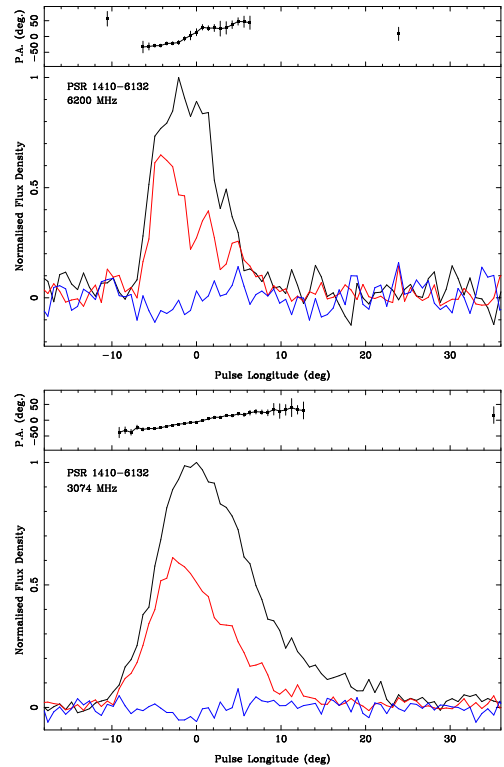

\begin{tabular}{c}
\centerline{\psfig{file=fig2a.ps,height=5cm,angle=-90}} \\
\centerline{\psfig{file=fig2b.ps,height=5cm,angle=-90}} \\
\end{tabular}
\caption{Polarisation profiles of PSR~J1410$-$6132 at 6.7~GHz (top)
and 3.1~GHz (bottom). The total intensity (black), linear polarisation
(red) and circular polarisation (green) traces are shown in the lower
panel of each plot. The upper panels show the position angles of the
linear polarisation.}
\label{fig:pol}
\end{figure}

\begin{figure}
\centerline{\psfig{file=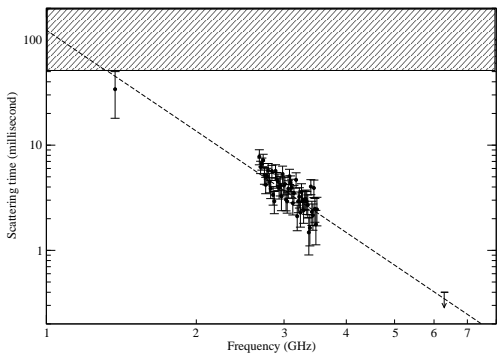,height=6cm,angle=-90}}
\caption{Measurement of the scattering broadening time of the pulses of
  PSR~J1410$-$6132 as a function of frequency at 1.4 GHz and across the
  3.1~GHz band and at 6.3 GHz. The derived scattering power law index is
  $-3.2\pm0.5$. The shaded area marks the region where the scatter broadening
  time exceeds the pulse period, preventing a detection of the pulsar.}
\label{fig:scattail}
\end{figure}

Figure~\ref{fig:pol} shows the polarisation profiles of the pulsar at 3.1 and
6.2~GHz. The first point to notice is the scatter broadening at the lower
frequency and indeed at 1.4~GHz (not shown) the profile is exceedingly
broadened by interstellar scattering. Under the assumption of Kolmogorov
turbulence, the time constant of the scatter broadening should vary with
frequency like $\nu^\beta$ with a scattering power law index
$\beta=-4.4$. Following L\"ohmer et al.~(2001), fits were made to the
scattering tails of the pulses of PSR~J1410$-$6132 as a function of frequency,
at 1.4 GHz and across the 3.1~GHz band. The result shown in Figure
\ref{fig:scattail} leads to an scattering power law index of $\beta =
-3.2\pm0.5$, significantly flatter than expected but in line with values seen
in other high DM pulsars (L\"ohmer et al. 2004).  We only obtain an upper
limit for the seemingly unscattered profile at 6.2~GHz which is rather narrow
with a full width of only 14\degr.  The linear polarisation is relatively
high, especially across the leading part of the profile. The position angle of
the linear polarisation swings through $\sim$80\degr\ across the profile.  The
narrow pulse width and the relatively flat gradient of position angle swing
preclude any determination of the geometry of the star.

\textbf{The flux densities at 1.5, 3.0 and 6.5~GHz are listed in Table~1.
At 1.5~GHz, the large scattering tail leads to a significant fraction
of the flux density being present as a DC term which is normally removed
during baseline subtraction. We have therefore corrected the measured
flux density of 4~mJy to 6~mJy to compensate for this effect.}
The flux density spectrum, $S\propto \nu^\alpha$, at
1.5, 3.0 and 6.3 GHz, has a power law index of $\alpha = -1.3\pm0.2$ 
which is somewhat flatter than the mean power law index of $-1.7$ 
derived from a large sample of pulsars (Maron et al.~2000).\nocite{mkk+00} 

\section{Discussion}

PSR~J1410$-$6132 is a young, highly energetic pulsar and it is located within
the positional error box of a previously unidentified $\gamma$-ray point
source, 3EG J1410$-$6147.  There are currently over 100 unidentified
$\gamma$-ray sources resulting from the $\gamma$-ray sky survey with the
Energetic Gamma Ray Experiment Telescope (EGRET), aboard the Compton Gamma-Ray
Observatory, which was performed from 1991 until 1999 in the energy range 100
MeV to 10 GeV \cite{hbb+99}. The nature of these unidentified point sources
has been debated for some time and many attempts have been made to find
corresponding counterparts at other parts of the electromagnetic
spectrum. Those that have been identified include solar flares, $\gamma$-ray
bursts, normal galaxies, active galactic nuclei and pulsars.

Pulsars are of particular interest because they have a similar 
spatial distribution as the unidentified EGRET point sources and have
been positively identified as being $\gamma$-ray emitters, e.g. the Crab and
Vela pulsars. Indeed, emission theories for pulsars predict $\gamma$-ray
emission in particular for young pulsars in either so called ``outer gap'' or
``polar cap'' models (see e.g.~Harding 2005 for a recent
review).\nocite{har05} In addition to seven pulsars clearly identified as
$\gamma$-ray sources due to their detected {\em pulsed} emission, several more
possible pulsar associations have been made with EGRET sources, often
motivated by positional coincidences (e.g.~Camilo et al. 2001; Kramer et
al. 2003\nocite{cbm+01,kbm+03}).  However, due to the large positional errors
of most $\gamma$-ray sources (of the order of 1$^{\circ}$), it is often
difficult to be certain if the association is real and final confirmation may
need the detection of corresponding pulsed emission using the new $\gamma$-ray
satellites like AGILE or the Gamma-Ray Large Area Space Telescope (GLAST).

\begin{table*}
\caption{Comparison of the parameters of the three possible 
pulsar associations with the EGRET source 3EG J1410$-$6147 
and the seven confirmed $\gamma$-ray pulsars.
Column 1 shows the pulsar name and EGRET source.
The variability index ($V$) according to McLaughlin (2001) is 
shown in column 2 with the distance to the pulsar in column 3.
This distance is based on the electron density model by Cordes \& Lazio
(2002) unless other information was available.
The offset between the pulsar position and the EGRET centroid is listed in
column 4, \textbf{where $\Delta\Phi$ is the separation between the EGRET source and the pulsar listed in column 1 and $\Theta_{95}$ is the size of the 95 per cent error-box}.
Column 5 gives the pulsar's $\dot{E}$ and the spin-down
flux $\dot{E}/d^{2}$ is shown in column 6. The efficiency $\eta$
of the conversion of
spin-down energy to $\gamma$-ray flux measured in the band 100 MeV--10 GeV
is given in the final column. The parameters were derived using
standard equations (see Lorimer \& Kramer 2005 for details).}
\begin{center}
\begin{tabular}{lllclcc}
\hline
Pulsar/3EG Source & \multicolumn{1}{c}{$V$} & \multicolumn{1}{c}{$d$} &
$\Delta\Phi$$/$$\Theta_{95}$ & \multicolumn{1}{c}{$\dot{E}$} &
\multicolumn{1}{c}{$\dot{E}$/{$d^2$}} & \multicolumn{1}{c}{$\eta$} \\
&   & (kpc) & & (erg~s$^{-1}$)  & (erg~s$^{-1}$kpc$^{-2}$) &
(per cent) \\
\hline
\noalign{\smallskip}
J1410$-$6132/J1410$-$6147   & 0.72 & 15.3  & 0.14    &   $1.0\times 10^{37}$   & $4.3\times 10^{34}$ & 10   \\ 
J1412$-$6145/$\quad ...$   & 0.72 & 7.8   & 0.41    &   $1.2\times 10^{35}$   & $1.4\times 10^{33}$  & 230     \\
J1413$-$6141/$\quad ...$   & 0.72 & 10.1  & 0.75    &   $5.6\times 10^{35}$   & $1.6\times 10^{33}$  & 84      \\
\noalign{\smallskip}
Crab/J0534$+$2200           & 0.50 & 2.0  &  1.23    &   $4.6\times 10^{38}$   & $1.2\times 10^{38}$ & 0.02    \\
\noalign{\smallskip}
Vela/J0835$-$4511           & 0.01 & 0.3  & 3.72   &   $6.9\times 10^{36}$   & $8.2\times 10^{37}$  & 0.07    \\
\noalign{\smallskip}
Geminga/J0633$+$1751        & 0.59 & 0.2  &  $-$    &   $3.2\times 10^{34}$   & $1.3\times 10^{36}$ & 2       \\
\noalign{\smallskip}
B1055$-$52/J1058-5234       & 0.04 & 0.7  & 0.66   &   $3.0\times 10^{34}$   & $5.8\times 10^{34}$ & 4 \\
\noalign{\smallskip}
B1509$-$58/(2EG 1443$-$6040)$^\ast$     &  0.99 & 4.2 &    $-$  &   $1.8\times 10^{37}$   & $1.0\times 10^{36}$  &   0.8  \\
\noalign{\smallskip}
B1706$-$44/J1710-4439       & 0.17 & 2.3  &  2.26  &   $3.4\times 10^{36}$   & $6.4\times 10^{35}$ & 1       \\
\noalign{\smallskip}
B1951$+$32/$^\dagger$       &  ?? & 3.2 &   $-$   &   $3.7\times 10^{36}$   & $3.6\times 10^{35}$ & 0.6  \\
\noalign{\smallskip}
\hline
\end{tabular}

{\footnotesize $^\ast$ COMPTEL detection. The weak 2EG source 
J1443$-$6040 associated by Kuipers et al.(1999) does not appear in 
the 3EG catalogue.\\
$^\dagger$ While clearly detected as pulsed $\gamma$-ray emission by
Ramanamurthy et al.~(1995), no associated sources appears in the 3EG catalogue.
}
\end{center}
\end{table*}
\nocite{mcl01,khk+99,rbd+95}

The true nature of EGRET source 3EG J1410$-$6147 has been long debated
in the literature \cite{sd95,cb99,djg+03,kbm+03}. 
The error box of 3EG J1410$-$6147 contains the supernova remnant (SNR)
G312.4$-$0.4 and two young energetic pulsars,
PSRs J1412$-$6145 and J1413$-$6141. It is possible that the SNR itself
is responsible for the $\gamma$-rays \cite{cb99} but the two pulsars
almost certainly are not, unless they have unusually high $\gamma$-ray
efficiencies or grossly underestimated distances \cite{djg+03,kbm+03}.
Doherty et al. (2003)\nocite{djg+03} speculated that the most likely
source powering 3EG J1410$-$6147 was a still hidden energetic pulsar.  We will
show here that our newly discovered pulsar is a very strong candidate for 
the long sought after counterpart of this well studied EGRET source.

Table 2 shows a comparison of the properties of the three energetic pulsars
coincident with the error box of 3EG J1410$-$6147, along with the seven
confirmed $\gamma$-ray pulsars. The variability index $V$ for each is shown in
column 2 (McLaughlin et al. 1996) and is $V<1$ for all pulsars and proposed
pulsar associations, indicative of non-variable flux. 3EG J1410$-$6147 has a
value of 0.72, highly indicative of a pulsar. \textbf{Column 4 shows the pulsar's position relative to the size of the EGRET 95\% confidence error ellipse}. The newly discovered pulsar is \textbf{the closest to the centroid of the error ellipse of 3EG J1410$-$6147 than the other two putative candidates.} The
spin-down luminosity $\dot{E}$ is shown in column 5 and the ``spin-down flux''
$\dot{E}$/{$d^2$}, which generally gives an indication of the detectability of
$\gamma$-ray pulsars, is shown in column 6. The observed efficiency for the
conversion of spin-down luminosity ($\dot{E}$) into $\gamma$-ray luminosity
($L_{\gamma}$) is shown in column 7. This value, $\eta$, was derived as in
Kramer et al.~(2003), using: $\eta$ = $L_{\gamma}$/$\dot{E}$ =
$\textit{f}$4$\pi$$d^{2}$$\bar{F}$/$\dot{E}$, where $\bar{F}$ is the observed
$\gamma$-ray flux and $\textit{f}$ is the $\gamma$-ray beaming fraction which
we assumed to be $\textit{f}$ $\equiv$ 1/4$\pi$.

The values in Table 2 show it very unlikely, from an efficiency point of view,
that either PSR J1412$-$6145 or J1413$-$6141 are associated with the EGRET
source unless their distances are overestimated by a factor of at least 3.  In
contrast, PSR J1410$-$6132 even at a distance of 15~kpc has an efficiency
similar to that of the known $\gamma$-ray pulsars and is located much closer
to the centroid of the $\gamma$-ray position than the other two pulsars within
the error box. The efficiency would be identical to that of PSR B1055$-$52 if
the real distance of PSR J1410$-$6132 were about 50\% less than that predicted
from the value of the DM. \textbf{Conversely, we could place the pulsar at a distance which is 50\% further than that derived from its DM and it would still have an efficiency no greater than the upper limit of 20\% considered plausible by Torres et al. (2001)}. \nocite{tbc01} Such uncertainties in distances based on the DM are not uncommon (Kramer et al.~2003), \textbf{and we therefore} claim that the new pulsar is the likely source of the $\gamma$-rays in 3EG J1410$-$6147.

Searching seven viewing periods of archival EGRET data, we performed an
analysis to detect pulsed $\gamma$-ray emission from the source, which was
limited by the low statistics and also possible timing noise and glitches. In
addition, large time intervals between viewing periods rendered optimal $\nu$
and $\dot{\nu}$ pairs for one viewing period inappropriate for studying
others. No evidence for a signal with a fluctuation detection probability
below 1\% was found in the seven data sets. However, with improved
sensitivity, observations with AGILE and GLAST will provide the opportunity to
search for periodic $\gamma$-ray emission with contemporaneous radio
ephemerides to confirm our claim.

The detection of this pulsar, in a region of the Galactic plane surveyed many
times before, re-opens the question as to the nature of the unidentified
sources found by EGRET in the plane. The low variability index of many sources
is typical of the known pulsar/EGRET source associations. It seems possible
therefore, that a high frequency search for pulsars in the EGRET error boxes
could uncover more distant, highly scattered pulsars previously missed in low
frequency surveys.  Such a survey would be timely prior to the launch of GLAST.

\section{Conclusions}
A survey with the Parkes telescope at the high frequency of 6.3~GHz
has been carried out along a thin strip of the Galactic plane.
We report here on the discovery of a young, highly energetic pulsar
located in the error box of the $\gamma$-ray source 3EG J1410$-$6147.
The parameters of the pulsar indicate that the association is highly likely,
although confirmation awaits the reduction of the $\gamma-$error box and/or
the detection of pulsations with AGILE or GLAST.

\section*{Acknowledgements}
This research was partly funded by grants from the Australian Research Council
and the Science \& Technology Facilities Council, UK. The Australia Telescope
is funded by the Commonwealth of Australia for operation as a National
Facility managed by the CSIRO. We thank A. Corogniu, R. Eatough and
R. Smits for helping with the observations and to M. Keith for helping to set
up the software applications. We are grateful to D.{~A.} Smith for useful
discussions.

%\bibliography{modrefs,psrrefs,crossrefs}
%\bibliographystyle{mn}

\label{lastpage}
\end{document}